\documentclass[reprint,aps,twocolumn,,floatfix,superscriptaddress,showkeys,showpacs]{revtex4-1}
\usepackage{graphicx}
\usepackage{physics}
\usepackage{commath}
\usepackage{bm}
\usepackage{dsfont}
\usepackage[usenames,dvipsnames]{xcolor}
\usepackage{pstricks}
\usepackage[tight]{subfigure}
\usepackage{verbatim}
\usepackage{units}
\usepackage{multirow}
\usepackage{enumitem}
\usepackage{mathrsfs}
\usepackage{leftidx}
\usepackage{xspace}
\usepackage{braket}
\usepackage{bbm}
\usepackage{amsfonts,amssymb,amsmath}
\usepackage{psfrag}
\usepackage[abs]{overpic}
\usepackage{float}
\usepackage[a4paper]{hyperref}
\hypersetup{colorlinks=true,linktoc=all,linkcolor=blue,breaklinks=true,citecolor=blue,urlcolor=blue}
\usepackage{lineno}  

\providecommand{\keywords}[1]
{
  \small	
  \textbf{\textit{Keywords---}} #1
}

\newcommand{\tongji}{Center for Phononics and Thermal Energy Science, China-EU Joint Lab on Nanophononics, Shanghai Key Laboratory of Special Artificial Microstructure Materials and Technology, School of Physics Science and Engineering, Tongji University, Shanghai 200092, China}

\newcommand{\znu}{Department of Physics, Zhejiang Normal University, Jinhua 321004, Zhejiang, China}

\begin{document}


\title{Geometric Heat Pump: Controlling Thermal Transport with Time-dependent Modulations}

\author{Zi Wang}
\author{Luqin Wang}
\author{Jiangzhi Chen}
\affiliation{\tongji}

\author{Chen Wang}
\email{Email: wangchen@zjnu.cn}
\affiliation{\znu}

\author{Jie Ren}
\email{Email: Xonics@tongji.edu.cn}
\affiliation{\tongji}

\date{\today}

\begin{abstract}
 The second law of thermodynamics dictates that heat simultaneously flows from the hot to cold bath on average. To go beyond this picture, a range of works in the past decade show that, other than the average dynamical heat flux determined by instantaneous thermal bias, a non-trivial flux contribution of intrinsic geometric origin is generally present in temporally driven systems.  This additional heat flux provides a free lunch for the pumped heat and could even drive heat against the bias. We review here the emergence and development of this so called ``geometric heat pump'', originating from the topological geometric phase effect, and cover various quantum and classical transport systems with different internal dynamics. The generalization from the adiabatic to the non-adiabatic regime and the application of control theory are also discussed. Then, we briefly discuss the symmetry restriction  on the heat pump effect, such as duality, supersymmetry and time-reversal  symmetry. Finally, we examine open problems concerning the geometric heat pump process and elucidate their prospective significance in devising thermal machines with high performance.


\end{abstract}
\maketitle{}
\tableofcontents

\section{Introduction}
According to the second law of thermodynamics, a system in steady state generally exhibits a directional flow on average driven by the thermodynamic bias, incurring an increase of entropy. This statement is universal regardless of the particular nature that the flow bears. For instance, particle/heat flows along the descending direction of the concentration/temperature in the steady state. Relaxing the restriction on the two-terminal flow direction requires breaking both the equilibrium and non-equilibrium steady state(NESS) conditions. This can be readily achieved by involving the time-dependent parametric driving. The effect of parametric driving on transport behaviors was originally unraveled in electron~\cite{brouwer1998scattering} and Brownian particle systems~\cite{hanggi2009artificial}. In electron pump studies, Coulomb blockade effect was found to be an essential ingredient in the generation of directed current in the adiabatic regime~\cite{aleiner1998adiabatic}. Independently, the flashing ratchet model was proposed to study the Brownian motor phenomena~\cite{hanggi2009artificial}. As the driving process is away from the adiabatic limit by having a finite frequency, intriguing results can be established, such as the photon-assisted tunneling~\cite{oosterkamp1997photon}, the coherent destruction of tunneling~\cite{grossmann1991coherent}, and non-adiabatic pumping~\cite{rahav2008directed, braun2008nonadiabatic, cavaliere2009nonadiabatic}, etc. There are also universal restrictions and no-pumping theorems on the pump effect, regardless of the driving frequency and amplitude~\cite{chernyak2008pumping, ren2011duality, asban2014no}. 

In parallel, the profound ideas of geometric phase~\cite{berry1984quantal}, also termed Berry phase, brought forth a surge of research aiming at defining gauge-invariant geometric observables and elucidating their effect in time-dependent systems, with Thouless pump~\cite{thouless1983quantization, nakajima2016topological} being one of the most important examples. Thouless formulated the linear response current integrated along one adiabatic driving period in a geometric expression, which is topologically quantized when the system is in a non-trivial zero temperature insulating phase. The Thouless pump can be intuitively understood as the quantum version of the well known Archimedes’ pump. While the above geometric pump research only concerns closed systems with extended dimensions, fluctuation of transport properties can be safely ignored. Later, this paradigm was generalized to cover open mesoscopic systems in matter/energy exchange with macroscopic reservoirs. The full characteristics of the fluctuating transport, encoded in the corresponding cumulant generating function(CGF), can be interpreted as a phase in the Berry-Sinitsyn-Nemenman sense~\cite{sinitsyn2007universal, sinitsyn2009stochastic}. In this way, the total CGF splits into the dynamic and geometric phase-like parts, in parallel with the argument of Berry phase. This geometric viewpoint offers a systematic classification of the pump effect and works as the basis of further analyses.

\begin{figure}[h]
 \centering
 \includegraphics[width=0.40\textwidth]{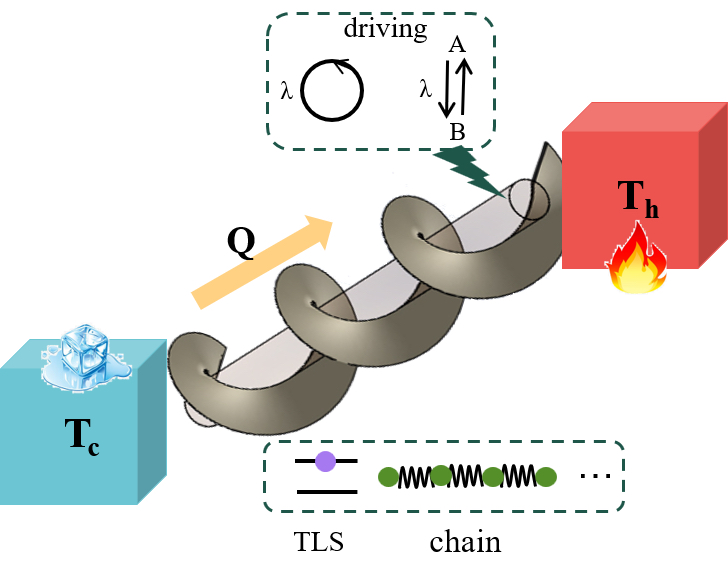}
 \caption{A schematic diagram of heat pump. The overall system is composed of two heat reservoirs and the middle system. A time-dependent driving protocol $\lambda(t)$ is introduced to pump heat flow from the cold reservoir to the hot reservoir, characterized as the temperatures $T_c$ and $T_h$, respectively. The driving parameter $\lambda(t)$ is either driven along a continuous contour or switched periodically between discrete values(A and B). The middle system can be a two-level system(TLS), a chain, etc. This figure is inspired by the Archimedes' screw. }
 \label{fig1}
 \end{figure}

The above works mainly concentrate on the geometric pump effect of charges and particles, without considering the thermodynamics of these processes. Nowadays, with the rapidly growing field termed stochastic thermodynamics, it is instrumental to explore the heat transport properties of mesoscopic/microscopic devices subject to a temporal driving, as shown in Fig.~\ref{fig1}, which pumps heat from a cold to a hot bath. Although the heat pump effect has been demonstrated in different systems~\cite{chamon2011heat, marathe2007two, segal2008stochastic, ren2010emergence}, an overall framework and systematic ways of studying are indispensable. Inspired by the geometric phase concept, it was pointed out in a series of research that the driving-induced heat pump has a universal geometric contribution~\cite{ren2010berry, ren2012geometric, chen2013dynamic}. The geometric heat pump has been shown to exist in anharmonic molecular junctions~\cite{ren2010berry}, classical oscillator systems~\cite{ren2012geometric}, quantum spin-boson systems~\cite{chen2013dynamic},etc. Considering the ubiquity of the geometric heat pump, it is demanding to further unveil the effect of various factors on its performance. In particular, the crossover of the system-bath interaction strength from weak to strong regimes~\cite{wang2017unifying} and the driving frequency from adiabatic to non-adiabatic regimes~\cite{ohkubo2008stochastic, uchiyama2014nonadiabatic} have been thoroughly studied. Also, the method developed in the search for shortcut to adiabaticity~\cite{guery2019shortcuts} and quantum control theory~\cite{d2007introduction} have been adapted to study the geometric pump effect~\cite{funo2020shortcuts, takahashi2020nonadiabatic}.

In this review, we mainly focus on the geometric heat pump effect in miscellaneous systems with different microscopic dynamics. In Section~\ref{sec2}, we review the theoretical framework of two-point measurement stochastic thermodynamics and the separation of CGF into the dynamic and geometric parts. In Section~\ref{sec3}, we make the general formalism specific by concentrating on the situation of quantum heat pump. The novel phenomenon of fractionally quantized heat pump is discussed. Also, we discuss the effect of the system-bath interaction strength on geometric pump. In Section~\ref{sec4}, we review the analogous effect found in classical systems. In Section~\ref{sec5}, we review the non-adiabatic geometric phase concept and show how the pump effect is modulated by the driving frequency. The application of control theory and shortcut methods is discussed. Furthermore, in Section~\ref{nogo}, the restriction on the pump effect by the symmetry arguments, such as duality, supersymmetry and time-reversal symmetry, is briefly illustrated. It is generally valid in the non-adiabatic regime, regardless of the driving protocol's detailed time dependence. Finally, in Section~\ref{sec6}, we summarize and try to point out further open questions. We mention here that, for the sake of clarity, the reduced Planck constant $\hbar$ and the Boltzmann constant $k_B$ are set to $1$ hereafter.

\section{General Formulation}
\label{sec2}

Whenever considering the mesoscopic dynamics, the noise with both the quantum and thermal nature is inevitable to study the transport properties of physical systems. Based on the well developed subjects termed counting statistics in both electron~\cite{kamenev2011field, levitov1993charge} and photon~\cite{kelley1964theory, gardiner2004quantum} communities, we can ask questions concerning the full probability distribution of the transported carrier number, which contains more information beyond linear response regime~\cite{andrieux2007fluctuation}. As an example, the full counting statistics(FCS) of electrons transported can shed light on the intrinsic coherent dynamics~\cite{stegmann2018coherent}. Recently, the FCS technique has been utilized to study the fluctuating heat/work in non-equilibrium systems, especially leading to a unified understanding of fluctuation theorems~\cite{esposito2009nonequilibrium, campisi2011colloquium}.

To study the FCS of heat transport, the essence is to assign a measure and the associated heat exchange to every possible trajectory~\cite{harris2007fluctuation}. This is transparent in the classical case, where the phase space trajectory concept is legitimate. Whereas, in the quantum regime, the trajectory is not well defined because of the uncertainty relations. Also, heat and work are not direct observables represented by Hermitian operators\cite{talkner2007fluctuation}. 

Nevertheless, the formalism of two-point measurement exactly resolves this issue~\cite{de2004quantum, esposito2006fluctuation, esposito2009nonequilibrium}. In this framework, the work done by the time-dependent driving can be always well defined as the total energy change of the overall system, while the definition of heat seems problematic in the strong coupling regime~\cite{talkner2020colloquium}. The controversy about the heat definition is settled if we concentrate on the weak coupling situation, where the heat exchange can be identified as the change of the reservoir energy. By introducing two projective measurements on the energy of reservoirs, we can define the characteristic function of the transported heat~\cite{esposito2009nonequilibrium}
\begin{equation}
\begin{split}
& Z(\chi, t)= \\
& \int dq_0 dq_t e^{\chi(q_t-q_0)}\textrm{tr}_S (\left | \braket{q_t|\hat{U}(t,0)|q_0} \right |^2 \braket{q_0|\hat{\rho}_0|q_0} ),
\end{split}
\end{equation}
where $\hat{U}(t,0)$ is the propagator from time 0 to t, fulfilling the Schr$\ddot{o}$dinger equation $i\frac{\partial}{\partial t}\hat{U}(t,0)=\hat{H}(t)\hat{U}(t,0)$($\hat{H}$ being the total system Hamiltonian). $\hat{\rho}_0$ is the initial density matrix. The notation $\textrm{tr}_S$ represents taking trace over the middle system's degrees of freedom. Here, two measurements on a heat bath are carried out before and after the temporal evolution from time $0$ to time $t$. The two-measurement results are denoted as $q_0$ and $q_t$, which are the transport quantities of prime interest observed at the initial and final measurements, respectively. $\ket{q_0}/\ket{q_t}$ is the corresponding energy eigenstate of the reservoir immediately after the initial/final measurement. In accordance with the quantum measurement postulate, $\left | \braket{q_t|\hat{U}(t,0)|q_0} \right|^2$ is the conditional probability of obtaining $q_t$ given the initial $q_0$. This result is generally valid, but may difficult to calculate directly. For convenient calculations, we make the assumption of the validity of Born-Markov approximation and the tensor product form of interactions. Accordingly, by taking the second-order perturbation with respect to the system-bath coupling strength in the interaction picture and tracing out the bath variables, the twisted quantum master equation can be derived. It has a form of Eq.~\ref{master equation} when written in the Liouville space using super-operators~\cite{silaev2014lindblad}. The main steps are re-derived in the appendix~\ref{appendix_quantum}. The corresponding derivation in classical systems is also given in the appendix ~\ref{appendix_classical}.

The heat transport problem is simplified under the above coarse-grained assumptions. The quantum and classical description attains a kind of similarity on this level. The evolution of distribution is described by the twisted Fokker-Planck/master equation as
\begin{equation}
\label{master equation}
\frac{\partial}{\partial t}\ket{\rho(\chi, t)}=\hat{L}(\chi, t)\ket{\rho(\chi,t)},
\end{equation}
where $\chi$ is the auxiliary counting field. Vector $\ket{\rho(\chi,t)}$ describes simultaneously the statistics of the reduced middle system and the transported heat. It relates to the twisted density matrix in the quantum case and the distribution function in the classical case. The operator $\hat{L}(\chi,t)$ governs the Markovian stochastic dynamics among different continuous/discrete states. In the situation with $\chi=0$ the equation is reduced to the traditional Fokker-Planck equation/master equation. The auxiliary parameter $\chi$ works as a convenient tool in obtaining the CGF of the transported heat. The way in which $\chi$ enters the equation is explicitly introduced in the appendix \ref{appendix_quantum} for the quantum case and \ref{appendix_classical} for the classical case. Notably, the Fokker-Planck and master equation apply to the diffusive and jump dynamics, respectively~\cite{gardiner1985handbook}.

In the temporally driven heat transport systems, the characteristic function of the accumulated heat is
\begin{equation}
Z(\chi)=\braket{-|\overleftarrow{\mathbb{T}} e^{\int_0 ^t dt' \hat{L}(\chi,t')}|\rho_0},
\end{equation}
where $\overleftarrow{\mathbb{T}}$ is the time-ordering operator. $\bra{-}$ is the normalized uniform distribution. $\ket{\rho_0}$ is the initial state of the middle system's reduced density matrix/probability distribution. An inner product with $\bra{-}$ amounts to making a summation/integral in the classical equations and to taking a trace in the quantum formulations.
The corresponding CGF is $G(\chi, t)=\ln Z(\chi, t)$. It generates all the $n$-th order cumulants of the accumulated transported heat by taking derivatives
\begin{equation}
 \left \langle Q^n \right \rangle_c=\frac{\partial^n G(\chi)}{\partial \chi^n} \big| _{\chi \to 0}.
\end{equation}
 
The cumulants introduced here are closely connected to the fluctuating thermal properties measurable in experiments. The first and second order cumulants quantify the mean heat flow $ \left \langle Q  \right \rangle_c \equiv \left \langle Q \right \rangle$ and the shot noise $\left \langle Q^2 \right \rangle_c \equiv \left \langle Q^2 \right \rangle-\left \langle Q \right \rangle^2$ in the thermal transport~\cite{larocque2020shot}, respectively. Here, we denote the $n$-th cumulants(moments) of one stochastic variable $X$ by $\left \langle X^n \right \rangle_c$( $\left \langle X^n \right \rangle$). Recently, the development of technology enables a direct extraction of single-shot statistics of the thermodynamic quantities from the experimental observations~\cite{maillet2019optimal}. Research along this line would further compare the theoretical calculation results concerning the ensemble averages with an average over a series of identical experimental realizations.

 \begin{figure}[h]
 \centering
 \includegraphics[width=0.4\textwidth]{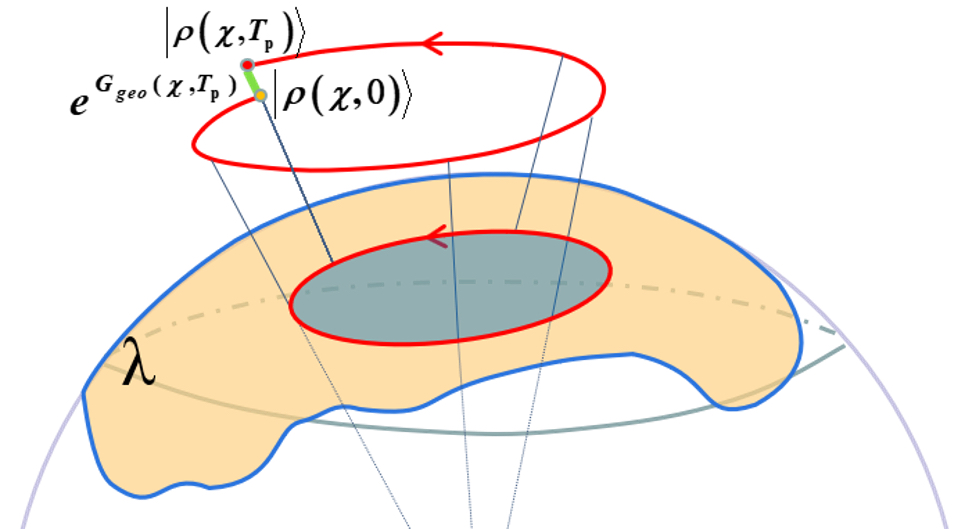}
 \caption{The schematic diagram of geometric phase in CGF. The twisted distribution on the parameter surface $\lambda$ at initial time is cyclically evolved into a vector parallel to itself, except for a non-trivial phase $G_{\textrm{geo}}(\chi, T_p)$. This phase has a geometric origin, since it only depends on the path traversed. }
 \label{fig2}
 \end{figure}

One constructive insight could be gained by observing that the CGF in driven systems is analogous to the phase accumulated by a quantum mechanical system under periodic driving~\cite{sinitsyn2007universal}, as shown schematically in Fig. \ref{fig2}. This idea originates from the Floquet generalization of Berry phase~\cite{aharonov1987phase}. The essence of this reasoning is briefly introduced in the appendix~\ref{appendix_geo}. Suppose now the distribution $\ket{\rho(\chi,t)}=e^{G(\chi, t)} \ket{\phi(\chi,t)}$, where $\ket{\phi(\chi,t+T_p)}=\ket{\phi(\chi,t)}$ is a cyclic solution of the master equation. The total CGF after one driving period is $G(\chi, T_p)$.
If a dual cyclic bra vector $\bra{\tilde{\phi}(\chi,t)}$ satisfying $\braket{\tilde{\phi}(\chi,t)| \phi(\chi,t)}=1$ exists, $G(\chi, T_p)$ is separated into the dynamical and geometric contributions $G(\chi, T_p)=G_{\textrm{dyn}}+G_{\textrm{geo}}$. These two CGF's are respectively
\begin{equation}
\label{phase split}
\begin{split}
    & G_{\textrm{dyn}}=\int_0 ^{T_p} dt{\braket{\tilde{\phi}(\chi, t)|\hat{L}(\chi, t)|\phi(\chi, t)}}, \\
    & G_{\textrm{geo}}=-\int_0 ^{T_p} dt{\braket{\tilde{\phi}(\chi, t)|\frac{\partial}{\partial t}|\phi(\chi, t)}}.
\end{split}
\end{equation}
Through a simple variable substitution, the geometric contribution can be equivalently written in a form depending on the driven parameter contour from $\vec{\lambda}(0)$ at time $0$ to $\vec{\lambda}(T_p)$ at time $T_p$:
\begin{equation}
\label{lambda}
    G_{\textrm{geo}}=-\int_{\vec{\lambda}(0)}^{\vec{\lambda}(T_p)} \braket{\tilde{\phi}(\chi,\lambda)|\frac{\partial}{\partial \vec{\lambda}}|\phi(\chi,\lambda)}\cdot d\vec{\lambda}.
\end{equation}
Here the vector $\vec{\lambda}$ is composed of the set of driven system parameters. This formula manifests the geometric origin of this component. The vector $\vec{A}=\braket{\tilde{\phi}(\chi,\vec{\lambda})|\frac{\partial}{\partial \vec{\lambda}}|\phi(\chi,\vec{\lambda})}$ is termed the geometric connection~\cite{berry1984quantal,aharonov1987phase}. The Stokes formula renders the expression $G_{\textrm{geo}}=-\int_\Omega \vec{F} \cdot d\vec{\Omega}$, where  $\vec{F}=\nabla\times \vec{A}$ is of the geometric nature of curvature, and the integration is over the parameter surface $\Omega$ enveloped by the parameter contour and $\vec{\Omega}$ is the unit normal vector on $\Omega$. One can easily check that the geometric phase contribution of CGF in the cyclic driving condition has a kind of ``gauge''-like invariance, i.e., when performing the ``gauge'' transformation
\begin{equation}
\begin{split}
    & \ket{\phi'(\chi, t)}=e^{f(\chi, t)}\ket{\phi(\chi, t)} \\
     & \bra{\tilde{\phi}'(\chi,t)}=e^{-f(\chi, t)}\bra{\tilde{\phi}(\chi,t)},
\end{split}
\end{equation}
where $f(\chi, t)$ is a single valued periodic function of $t$. 

The general expression of dynamic and geometric CGF, Eq.~\ref{phase split}, works in both adiabatic and non-adiabatic regimes. And it provides an equal treatment of both quantum and classical systems. Specifically, the dynamical and geometric parts can be naturally distinguished by their different scaling behaviors with respect to the driving period in the adiabatic regime, in parallel with the Berry phase~\cite{berry1984quantal}. Obviously, the dynamical part is of the same order as $T_p$. In contrast, the geometric part is independent of the length of $T_p$ (also, the driving velocity) as long as the adiabatic condition is satisfied. This certifies the concept of the geometric heat flux. Also, the geometrically transported heat has no steady state counterpart, since the geometric contribution to the CGF is zero when no driving is applied. Strictly speaking, in the non-adiabatic regime, the geometric part not only depends the length of time period, but also the detailed driving protocol, like the situation in non-adiabatic Aharonov-Anandan(AA) phase~\cite{aharonov1987phase}. Nevertheless, the term ``geometric" is preserved at non-adiabatic cases due to the expression of curvature. The only difference is that the curvature is defined on the state space that is locally defined 
by the Floquet state, dependent on the traversed driving parameters. When driving is very slow, the Floquet state reduces to the instantaneous eigenstate that generates the traditional adiabatic geometric curvature. 

For later convenience, we emphasize that although generally the cyclic bras $\bra{\tilde{\phi}(\chi, t)}$ and kets $\ket{\phi(\chi,t)}$ are left for the treatment of the Floquet method, they can be universally approximated by the instantaneous steady states if we are interested in the adiabatic regime. We denote the steady states by $\bra{l_0}$ and $\ket{r_0}$ in the following discussion, with the subscript $0$ denoting the steady state. They correspond to the right and left steady state, specified by the eigenvalue equation $\hat{L}\ket{r_0}=E_0 \ket{r_0}$ and $\bra{l_0}\hat{L}=\bra{l_0}E_0$, where $E_0$ is the eigenvalue with the largest real component.

The above formulation is restricted to the driving with continuous protocol, which can be also adapted to discrete switching models, in a similar way of Berry phase's definition~\cite{resta2011insulating}. The geometric CGF induced by the instantaneously switching system parameter from $\lambda(A)$ to $\lambda(B)$ reads
\begin{equation}
\label{dis_geo}
G_{\textrm{geo}}=\ln \braket{l_0(\chi, \lambda(B))|r_0(\chi, \lambda(A))}.
\end{equation}
 Since in open systems the operator $\hat{L}$ is generally non-Hermitian, the geometric phase contribution by cyclically switching between $A$ and $B$ is non-trivial: $G_{\textrm{geo}}=\ln(\braket{l_0(\chi, \lambda(B))|r_0(\chi, \lambda(A))}\braket{l_0(\chi, \lambda(A)) | r_0(\chi, \lambda(B))})\neq0$. This effect has been utilized in constructing thermal-electric conversion machine~\cite{ren2014third}.

We note that the discrete geometric phase effect is naturally related to the continuous one~\cite{resta2011insulating}. By taking the infinitesimal protocol variation $\Delta\lambda=\lambda(A)-\lambda(B)$ limit and neglecting higher-order terms, $\ket{r_0(\chi,\lambda(A))}=\ket{r_0(\chi,\lambda(B))}+\Delta\lambda\frac{\partial}{\partial \lambda}\ket{r_0(\chi,\lambda)}|_{\lambda=\lambda(B)}$. The evaluated geometric phase $G_{\textrm{geo}}$ in Eq.~\ref{dis_geo} is up to first order $G_{\textrm{geo}}=\Delta\lambda\braket{l_0(\chi,\lambda)|\frac{\partial}{\partial \lambda}r_0(\chi,\lambda)}|_{\lambda=\lambda(B)}$. By performing a cyclic loop integral, we 
achieve the adiabatic counterpart of the same form of the geometric CGF given by Eq.~\ref{lambda}.

\section{Quantum Geometric Heat Pump}
\label{sec3}
The fascinating advancement of nano-technology has been largely stimulated by the ever-growing desire for more powerful functional devices. The miniaturization of devices inevitably renders the underlying quantum nature more pronounced than ever before. The analysis of their thermodynamics remains a challenging and rewarding task. The rewards are represented by quantum engines controlling and harnessing the thermal energy with a high performance~\cite{li2012colloquium, josefsson2018quantum, seah2020maxwell, abiuso2020optimal, marcos2018thermal}.

A general twisted quantum master equation given in appendix~\ref{appendix_quantum} simultaneously governs the evolution of the dynamics and the counting statistics of heat transport. It is valid as long as the relatively weak coupling condition between the middle system and the baths is assumed. It works as an efficient technique in describing non-harmonic transport phenomena, in contrast to the non-equilibrium Green's function approach~\cite{kamenev2011field}. Recent research on quantum geometric heat pump has considered the molecular junction system~\cite{ren2010berry}, the spin-boson systems~\cite{chen2013dynamic, wang2017unifying}, and the optomechanical systems~\cite{nie2020berry} using the twisted master equation formalism in various basis. In the paper~\cite{nie2020berry}, the correspondence between the quantum master equation and a generalized Fokker-Planck equation in the coherent state basis is utilized.

The quantum master equation in super-operator Liouville space~\cite{silaev2014lindblad} is given by the linear equation Eq.~\ref{master equation}. The operator $\hat{L}(\chi, t)$ determines a set of instantaneous eigenvalues and eigenvectors by equations 
\begin{equation}
\label{eigen}
    \begin{split}
    & \hat{L}(\chi, \lambda)\ket{r_n(\chi, \lambda)}=E_n(\chi, \lambda)\ket{r_n(\chi, \lambda)} \\
    & \bra{l_n(\chi, \lambda)}\hat{L}(\chi, \lambda)=E_n(\chi,\lambda)\bra{l_n(\chi,\lambda)}. 
    \end{split}
\end{equation}
Here $\lambda$ is the system parameters implicitly time-dependent.

\begin{figure}[h]
 \centering
 \includegraphics[width=0.45\textwidth]{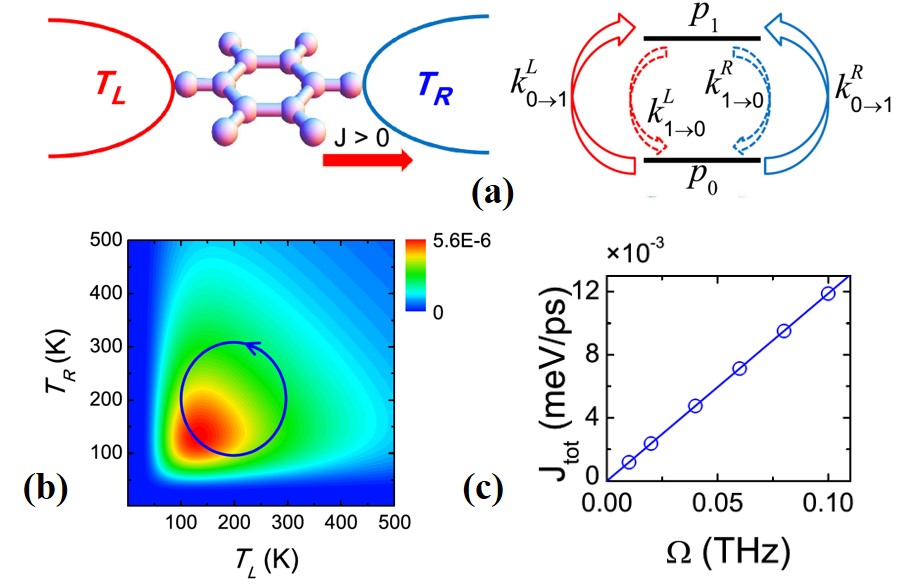}
 \caption{The geometric heat pump effect in a molecular junction system. (a) The system setup. The molecule can exchange heat with the two baths, via transitions between two energy levels. The transition rates($k$) correspond to different possible events. (b) The geometric structure of the open system on the parameter surface of left and right temperature values. The colorbar shows the Berry curvature of the local ground state. A geometric heat pump can be induced by encircling a non-trivial region. (c) The time averaged geometric part of the pumped heat versus the driving frequency. The linearity verifies its geometric origin. Figures are reproduced from \cite{ren2010berry}. }
 \label{fig3}
 \end{figure}

The quantum master equation in the energy-level basis has a simple form if the quantum coherence effect is negligible. This is the effective dynamics in the long time and weak coupling limit~\cite{wang2017unifying}. For a middle system with two energy levels, this twisted Pauli master equation has the twisted transition matrix~\cite{ren2010berry}
\begin{equation}
\hat{L}(\chi)=\begin{pmatrix}
-k_{0 \to 1}^L-k_{0 \to 1}^R & k_{1 \to 0}^L+k_{1 \to 0}^Re^{\chi} \\
k_{0 \to 1}^L+k_{0 \to 1}^R e^{-\chi} &-k_{1 \to 0}^L-k_{1 \to 0}^R,
\end{pmatrix}
\end{equation}
where the physical meaning is illustrated in Fig. \ref{fig3}. The subscripts $0$ and $1$ are the states of the middle system, while the superscripts $R$ and $L$ are the two transition channels. If the transition with $R$ happens, the energy of the middle system is transmitted into or absorbed from the right reservoir in an energy conserving manner. So the counting field $\chi$ counts the net number of system's spontaneous and stimulated transitions by exchanging energy quanta with the right reservoir. This gives us an intuitive sense of the physical process in full-counting the heat transport.

The eigenvalues of $\hat{L}(\chi)$ can be sorted in the decreasing order of the real part of the corresponding eigenvalues. ${E_n(\chi, t)}$ is the sorted result. The largest eigenvalue $E_0(\chi, t)$ gives the dominant CGF in the long-time limit, according to the large deviation theory~\cite{touchette2009large}. This accounts for the full fluctuation properties of heat transport in the steady states. $\Delta E(\chi, t)=E_0(\chi,t)-E_1(\chi, t)$ is the eigenvalue difference of the first two states. Its inverse $\frac{1}{\Delta E(\chi=0, t)}$ characterizes the time-scale of reaching to the steady states if the system parameters are fixed to the value $\lambda(t)$ at time $t$. If the time-scale of driving is much larger than $\frac{1}{\Delta E(\chi=0, t)}$, the adiabatic limit of the geometric heat pump is reached. In this regime, the time-dependent distribution is equal to the instantaneous steady states at each time. So the right and left eigenvectors $\ket{r_0(\chi, t)}$, $\bra{l_0(\chi, t)}$ define the natural cyclic basis(discussed in appendix~\ref{appendix_geo}). The CGF of pumped heat is separated as
\begin{equation}
\begin{split}
    & G_{\textrm{dyn}}=\int_{0}^{T_p} E_0(\chi, t) dt \\
     & G_{\textrm{geo}}=-\int_{\vec{\lambda}(0)}^{\vec{\lambda}(T_p)} \braket{l_0(\chi, \vec{\lambda}(t))|\frac{\partial}{\partial \vec{\lambda}}|r_0(\chi, \vec{\lambda}(t))} \cdot d \vec{\lambda}.
\end{split}
\end{equation}
The geometric part is expressed in a Berry-phase-like form. It is a line integral of the geometric connection $A_{\lambda_i}=\braket{l_0(\chi, \vec{\lambda})|\frac{\partial}{\partial \lambda_i}|r_0(\chi, \vec{\lambda})}$. Using the Stokes formula, when involving two parameters this can be identified as a surface integral of gauge-invariant Berry curvature contribution
\begin{equation}
\begin{split}
    G_{\textrm{geo}}&=-\int_\Omega d\lambda_1 d\lambda_2  F_{\lambda_1,\lambda_2}, \\
     \text{with} \;\;
    F_{\lambda_1,\lambda_2}&=\frac{\partial }{\partial \lambda_1}A_{\lambda_2}-\frac{\partial}{\partial \lambda_2}A_{\lambda_1},
\end{split}
\end{equation}
where $\Omega$ is the surface area encircled by the driving protocol in the parameter space and the curvature can be rewritten as $F_{\lambda_1,\lambda_2}=\braket{\frac{\partial}{\partial \lambda_1}l_0(\chi, \vec{\lambda})|\frac{\partial}{\partial \lambda_2} r_0(\chi, \vec{\lambda})}-\braket{\frac{\partial}{\partial \lambda_2}l_0(\chi, \vec{\lambda})|\frac{\partial}{\partial \lambda_1} r_0(\chi, \vec{\lambda})}$.

\begin{figure}[h]
 \centering
 \includegraphics[width=0.43\textwidth]{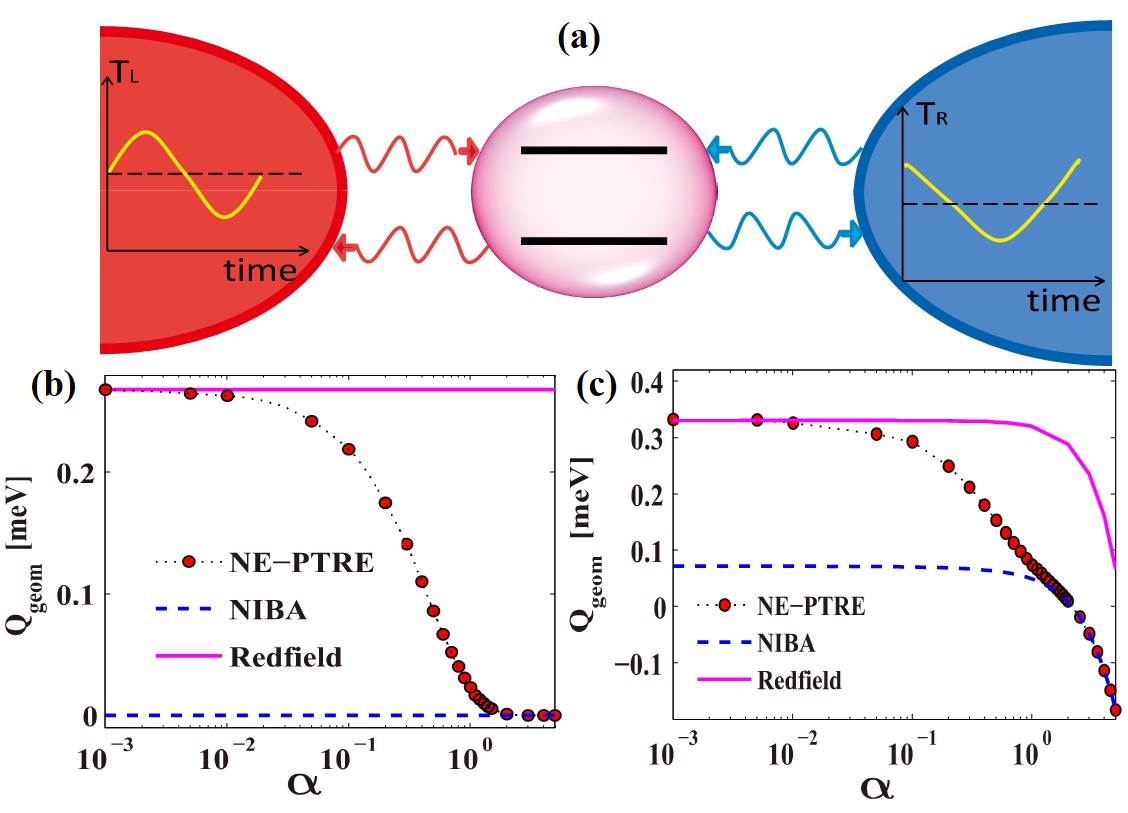}
 \caption{The dependence of geometric heat pump effect on the coupling strength $\alpha$ between the middle system and heat baths. As reviewed in the main text, the non-equilibrium polaron-transformed Redfield equation(NE-PTRE) method is valid in the crossover regime between weak(Redfield) and strong(NIBA) coupling limits. (a)A sketch of the driven spin-boson system. (b) When there is no Zeeman energy level splitting, the geometric heat pump effect vanishes as increasing the system-bath coupling strength. (c) When there is a given finite Zeeman term, the geometric heat pump effect is strong in the two limits of extremely weak and strong coupling, while the reversal of geometric heat pump effect occurs in an intermediate regime. Figures are reproduced from \cite{wang2017unifying}. }
 \label{fig4}
 \end{figure}

The research~\cite{ren2010berry} first showed that, by modulating temperatures of the two reservoirs $T_L$ and $T_R$, a non-trivial Berry curvature leads to a non-vanishing geometric heat pump phenomenon, as shown in Fig.~\ref{fig3}. Remarkably, the geometric contribution breaks the typical form of the steady state fluctuation theorem and encircling the whole parameter surface gives a novel fractional quantization of heat pump. Several following studies showed the similar effect in various quantum systems~\cite{chen2013dynamic, wang2017unifying, nie2020berry}. One of the most important discoveries along this line is the unveiling of the role of system-bath coupling strength in the geometric heat pump~\cite{wang2017unifying}. It demonstrated an intriguing behavior in the coupling's modulation effect. As illustrated in Fig.~\ref{fig4}, when there is no Zeeman energy splitting in the middle system, the geometric pump effect is gradually eliminated by increasing the system-bath coupling strength. But in presence of the energy splitting, $Q_{\textrm{geo}}$ has the largest absolute values and opposite signs in the weak and strong coupling limit, while crossing zero in an intermediate system-bath coupling regime, indicating a coupling-strength-induced geometric heat flux reversal. A full discussion of more diverse system parameters' influence is useful for future practical experiments and applications.

Also, the geometric pump effect can be transplanted into interacting systems with multi-degrees of transport carriers. As shown in the past research~\cite{ren2014third}, a novel thermal-electric conversion mechanism can be fueled by the geometric phase contribution in temporally driven systems. This work indicates a future profitable application of the geometry concept to devising energy harvesting machines.

\section{Classical Geometric Heat Pump}
\label{sec4}
The amplitude of the noise depends strongly on the temperature of heat baths. As the temperature is raised to a point where the thermal noise becomes dominant, the heat conduction process can be well described by a set of classical equations. 

Classical heat transport turns out to be sufficient in modeling a whole range of systems~\cite{dhar2008heat} and quite helpful in discovering devices with novel functions, like thermal diode, thermal memory, etc~\cite{li2012colloquium}. In addition to these steady state studies, it was found that the time-dependent driving in classical systems can be a convenient way of heat manipulation~\cite{marathe2007two, ren2010emergence,torrent2018nonreciprocal, segal2006molecular}. These research works demonstrated the existence of heat pump effect in simple models. Notably, a non-zero heat flow can be maintained by the driving at strict zero temperature bias~\cite{ren2010emergence}.

\begin{figure}[h]
 \centering
 \includegraphics[width=0.47\textwidth]{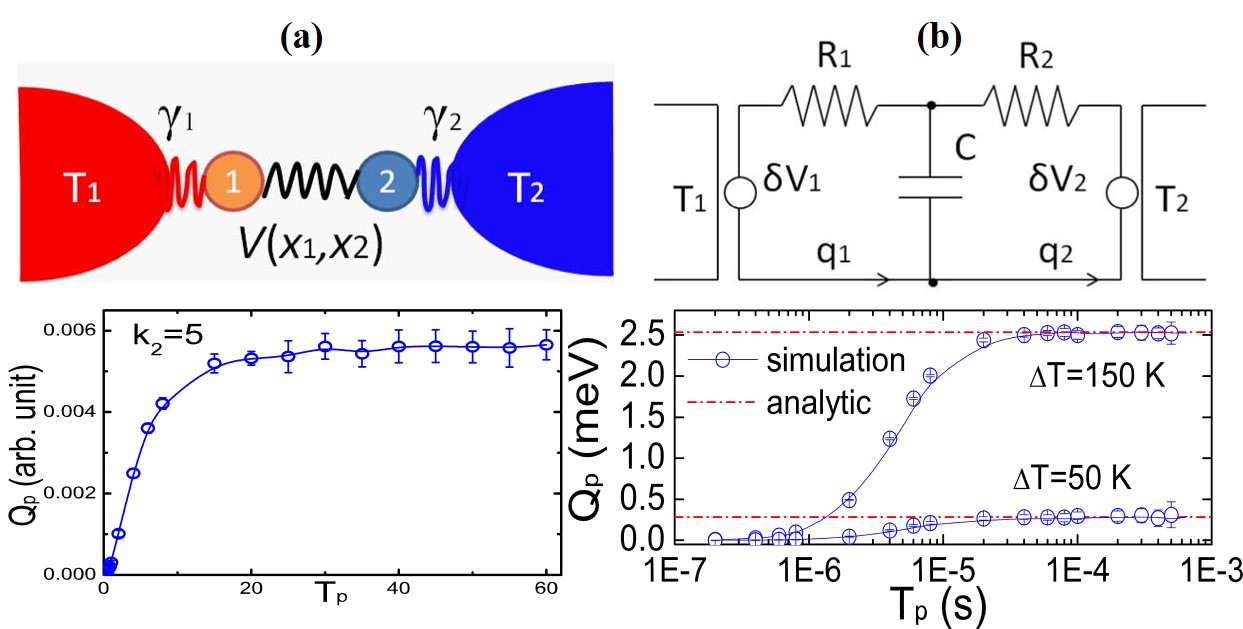}
 \caption{Two platforms of the classical geometric heat pump. Left panel: (a) a coupled Brownian oscillator system. The geometric pumped heat per driving time period is independent of the driving frequency in the adiabatic (large $T_p$) limit, which on the other hand approaches zero in the fast driving limit $T_p\rightarrow0$. Right panel: (b) a coupled classical electric circuit system. The geometric pumped heat shows similar behaviors. Here $\Delta T$ is the driving amplitude of time dependent temperatures. Figures are reproduced from~\cite{ren2012geometric}. }
 \label{fig5}
 \end{figure}

In order to study the heat pump phenomenon where more than one time-dependent parameter are involved, the concept of classical geometric heat pump was put forward~\cite{ren2012geometric, arrachea2012microscopic}. Using an analytically solvable classical model, the non-trivial geometric phase contribution was explicitly identified. In the coupled overdamped Langevin oscillator systems depicted in  Fig.~\ref{fig5}a, the equation of motion is~\cite{ren2012geometric}
\begin{equation}
\begin{split}
    & \gamma_1 \dot{x}_1+\frac{\partial}{\partial x_1}V(x_1,x_2)=\xi_1 \\
     & \gamma_2 \dot{x}_2+\frac{\partial}{\partial x_2}V(x_1,x_2)=\xi_2.
\end{split}
\end{equation}
$V(x_1,x_2)=\frac{1}{2}k(x_1-x_2)^2$ is the interaction potential between two oscillators. Following the general method and notation given in appendix ~\ref{appendix_classical}, the twisted Fokker-Planck equation is of the same form as Eq.~\ref{master equation}, with the twisted Fokker-Planck operator given by
\begin{equation}
\begin{split}
    & \hat{L}(\chi, t)=(\frac{T_1}{\gamma_1}+\frac{T_2}{\gamma_2})\frac{\partial ^2}{\partial y^2}+k(\frac{1}{\gamma_1}+\frac{1}{\gamma2}-\frac{2T_1}{\gamma_1}\chi)y\frac{\partial}{\partial y} \\
     & +k^2(\frac{T_1}{\gamma_1}\chi^2-\frac{\chi}{\gamma_1})y^2+k(\frac{1}{\gamma_1}+\frac{1}{\gamma_2}-\frac{T_1}{\gamma_1}\chi),
\end{split}
\end{equation}
where the variable $y=x_1-x_2$. Similarly to the quantum situation in Section~\ref{sec3}, the left and right eigenvectors of $\hat{L}(\chi,t)$ provide a natural local basis of cyclic state in the adiabatic regime. Denote the eigenvalues, bi-orthogonal left and right eigenvectors as $E_n(\chi, t)$, $\bra{l_n(\chi, t)}$ and $\ket{r_n(\chi, t)}$ respectively. $\hat{L}(\chi, t) \ket{r_n(\chi, t)}=E_n(\chi, t)\ket{r_n(\chi, t)}$ and $\bra{l_n(\chi, t)} \hat{L}(\chi, t)=E_n(\chi, t)\bra{l_n(\chi, t)}$. $n=0$ corresponds to the non-equilibrium steady state. As such, the geometric and dynamical components of the CGF phase are, respectively
\begin{equation}
\begin{split}
    & G_{\textrm{dyn}}=\int_{0}^{T_p}dt E_0(\chi, t) \\
     & G_{\textrm{geo}}=-\int_{0}^{T_p}dt \braket{l_0(\chi,t)|\frac{\partial}{\partial t}|r_0(\chi, t)}.
\end{split}
\end{equation}

Remarkably, the geometric part is independent of the driving period. It is uniquely determined by the geometric structure on the parameter space and the traversed contour. This extra contribution does not vanish in the infinite slow driving limit. Similar to the quantum situation, the modulation of two temperatures produces a non-trivial geometric pump~\cite{ren2012geometric}. If the dynamics of the middle system is non-linear, its effective parameters in the effective harmonic dynamics are temperature-dependent~\cite{li2007temperature}. This makes the application of the geometric heat pump effect more versatile. The direct generalization applies to the noisy RLC circuit~\cite{ren2012geometric} and the classical chain system~\cite{arrachea2012microscopic}. Studies concerning other classical systems is necessary for making contact with the current fabrication technology of thermal devices.

There are also a variety of studies demonstrating the heat pump effect without realizing the role of geometric phase effect plays in the process. As we concentrate on the geometric heat pump, we merely mention some of them, with no attempt at a thorough covering. The stochastic-resonance-like heat pump~\cite{ren2010emergence}, the passive phonon pump~\cite{chamon2011heat}, the heat pump approaching Carnot efficiency~\cite{segal2008stochastic}, and radiative heat pump effect ~\cite{li2019adiabatic}, as prominent examples, have been proposed and analyzed. It is foreseeable that the geometric concepts would instill further insights into these theoretical modeling~\cite{bhandari2020geometric} and the future experimental design of heat engines.

The length scale of the studied system is also an important factor. The well studied size effect in stationary nanophononic systems~\cite{li2012colloquium} has its analogy in the driven heat pump systems. As an example, by changing the system size, it is possible to reverse the direction or optimize the magnitude of the pumped heat flow in an anharmonic lattice system~\cite{ren2010emergence}. As the system  size is raised, the characteristic frequency of its collective dynamics will decrease. Therefore, for a given driving frequency, the change of system size will cause the change between adiabatic and non-adiabatic regime when comparing to the changing characteristic frequency.

Also, we would like to point out that the geometric heat pump effect in classical cases with Langevin baths is much more like that in quantum cases with the weak system-bath coupling limit, where driving system-bath couplings arbitrarily does not produce any non-trivial geometric pump effect~\cite{ren2010berry,ren2012geometric}. Since the strong system-bath coupling in quantum cases modifies the effective spectrum of heat baths, we speculate that in classical cases the different heat baths with different spectrum will have different impact on the geometric heat pump effect, which deserves further studies. Moreover, the effect of classical-to-quantum crossover on the geometric heat pump is also interesting to study in the future.


\section{Non-adiabatic Geometric Pump Effect}
\label{sec5}
The universal adiabatic limit of pumping will be broken if finite rate driving protocols are allowed. The effect can be understood as originating from transitions from the initial (steady, eigen) state to other possible states during the driving process. This fact is reminiscent of the celebrated Fermi golden rule, the Landau-Zener formula, etc, in conventional quantum mechanics~\cite{landau2013quantum}. Due to the non-adiabatic transitions from the instantaneous steady state, the non-adiabatic geometric pump effect may be not solely determined by the geometric structures of the time-independent parameter space. This can be manifested in the deviation from the adiabatic behavior. The pumped quantity is determined by the explicit form of the time dependence of driving protocols, aside from the contour of passed parameters \cite{ohkubo2008stochastic, ohkubo2010noncyclic, uchiyama2014nonadiabatic, takahashi2020nonadiabatic, funo2020shortcuts}.

Nevertheless, the separation of pumped heat into the geometric and dynamical parts is well defined using the Floquet cyclic state~\cite{ohkubo2008stochastic, ohkubo2010noncyclic}. The details are given in the appendix \ref{appendix_geo}. It is expected that the geometric contribution is connected to the excess entropy production, which may be an essential quantifier in non-equilibrium fluctuation-dissipation theorems~\cite{harada2005equality, lippiello2014nonequilibrium}.

\begin{figure}[h]
 \centering
 \includegraphics[width=0.45\textwidth]{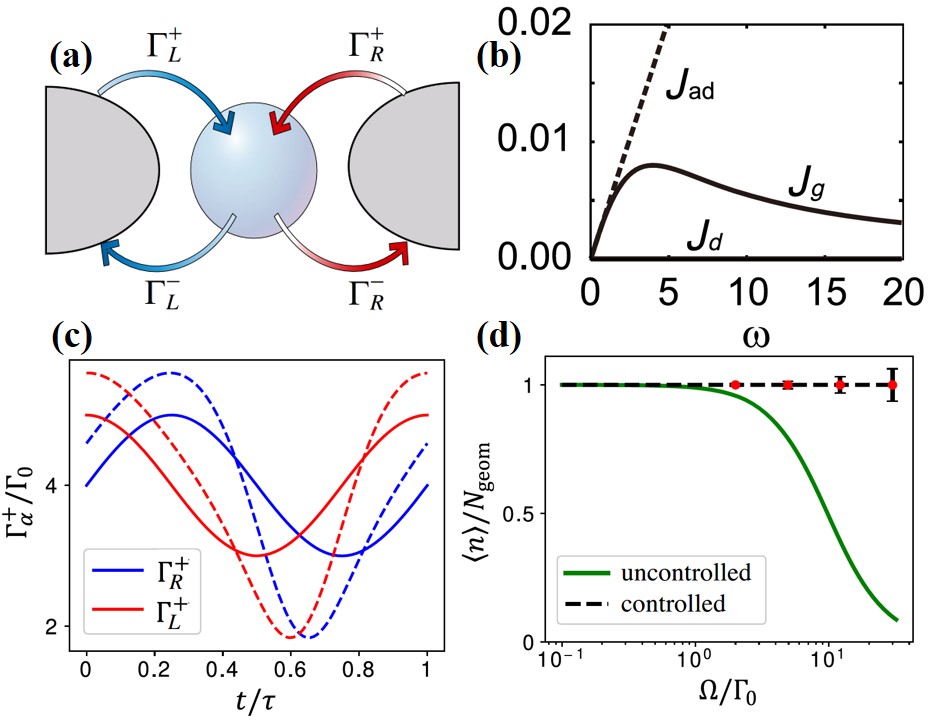}
 \caption{The non-adiabatic control of geometric pump effect. (a)The schematic setup of the stochastic pump. The superscripts of transition rates imply particle's entering/leaving the middle system, while the subscripts denote two transition channels(left/right). (b) The different components of pumped current versus the driving frequency. $J_d$, the dynamical component of pumped heat, is strictly zero in this setup. In this way, the geometric part $J_g$ would be the only present current. The adiabatic description $J_{ad}$ fails in the large frequency regime. (c) Controlling over the geometric pump. The solid line is for the uncontrolled driving protocol, while the dashed line represents the situation where the control is adopted. (d) The performance of non-adiabatic control. The geometric pump is enormously enhanced in the large frequency regime, by taking advantage of the control protocol. Figures are reproduced from \cite{takahashi2020nonadiabatic, funo2020shortcuts}. }
 \label{fig6}
 \end{figure}

The non-adiabatic geometric formulation is quite elegant, and works as a general framework for all driving frequencies. But the results generally depend on both the driving speed and the driving contour in a complicated way. It lacks an intuitive geometric picture. In addition, as shown in Fig.~\ref{fig6}d, the geometric part is gradually diminished in the large frequency (fast driving) regime, strongly restricting the power of the geometric pump as a practical way of charge/heat manipulation.

As shown by the recent research, the shortcut to adiabaticity method based on the control theory can be adopted in controlling the non-adiabatic geometric pump~\cite{takahashi2020nonadiabatic, funo2020shortcuts}. The original shortcut method mainly concentrates on achieving a fast and robust generation of a given state in closed systems~\cite{guery2019shortcuts}. Its guiding principle is adding an auxiliary pre-designed time-dependent Hamiltonian $\hat{H}_c(t)$ to the bare Hamiltonian $\hat{H}_0(t)$, so as to smoothly go from an eigenstate of $\hat{H}_0(0)$ to that of $\hat{H}_0(t)$~\cite{del2013shortcuts}.  In stead of rapidly preparing a final eigenstate of a certain Hamiltonian, the non-adiabatic control over the geometric pump robustly enhances the performance of fast pumping.

Here we briefly review the spirit of these works~\cite{takahashi2020nonadiabatic, funo2020shortcuts}. We first focus on the near-adiabatic regime, although the general generalization should be formulated by adopting the Floquet formalism. The general master/Fokker-Planck Eq.~\ref{master equation} can be comprehended as an eigenvalue equation $(\hat{L}( t)-\frac{\partial}{\partial t})\ket{\rho_1(t)}=0$. Near adiabatic limit, the operator $ -\frac{\partial}{\partial t}$ can be considered as a perturbation. The zeroth order(no time-dependent driving) solution of the perturbed equation is the instantaneous steady state $\ket{r_0(t)}$. The adiabatic perturbation theory~\cite{kolodrubetz2017geometry} formulates the first order correction as $\ket{\delta r_0(t)}=\hat{L}^{-1}\ket{\partial_t r_0(t)}$, where $\hat{L}^{-1}$ is the pseudo-inverse of $\hat{L}(t)$. Therefore, $\ket{\rho_1(t)}=\ket{r_0(t)}+\ket{\delta r_0(t)}$. Here, this geometric contribution can also be related to the adiabatic gauge potential~\cite{kolodrubetz2017geometry}. 

The non-adiabaticity of the driving induces the deviation from this first order approximation, since non-adiabatic driving $\hat{L}(t)$ generates distribution higher than the first order approximation valid approximately in the near adiabatic regime. To eliminate this deviation, an additional $\hat{L}_{cont}(t)$ can be devised to generate the equivalent distribution as in adiabatic driving. $\hat{L}+\hat{L}_{cont}$ acts as a generator of the evolution from time $t$ to $t+\delta t$($\delta t$ is infinitesimal):
\begin{equation}
\begin{split}
    & [1+(\hat{L}(t)+\hat{L}_{cont}(t))\delta t](\ket{r_0(t)}+\ket{\delta r_0(t)})\\
     & =\ket{r_0(t+\delta t)}+\ket{\delta r_0(t+\delta t)}.
\end{split}
\end{equation}
The specific form of $\hat{L}_{cont}$ is of no concern here, but it should be noted that the applied control extends enormously the geometric pump regime, as shown in Fig.~\ref{fig6}d. As long as the form of $\ket{\rho(t)}$ is preserved, the pumped heat in one driving period is
\begin{equation}
Q=\int_{0}^{T_p}dt (\braket{-|\hat{J}|r_0(t)}+\braket{-|\hat{J} \hat{L}^{-1}|\partial_t r_0(t)}).
\end{equation}
$\hat{J}$ is the current operator, which is normally independent of time. $\ket{-}$ is a uniform distribution. The contribution $\int_{0}^{T_p}dt \braket{-|\hat{J} \hat{L}^{-1}|\partial_t r_0(t)}=\int_{\lambda_0}^{\lambda_{T_p}}d \lambda \braket{-|\hat{J} \hat{L}^{-1}|\partial_\lambda r_0(\lambda)}$ is independent of the duration of the driving protocol $\lambda(t)$. Thus it is of the geometric origin, similar to the general remarks given in Section~\ref{sec2}. The existent studies concerning controls over the geometric pump have only been focused on the non-adiabatic control over the average current, with only the uncontrolled fluctuation of current discussed numerically~\cite{takahashi2020nonadiabatic}.

\section{Symmetry Constraint on Non-adiabatic Pump}
\label{nogo}

\begin{figure}[h]
 \centering
 \includegraphics[width=0.45\textwidth]{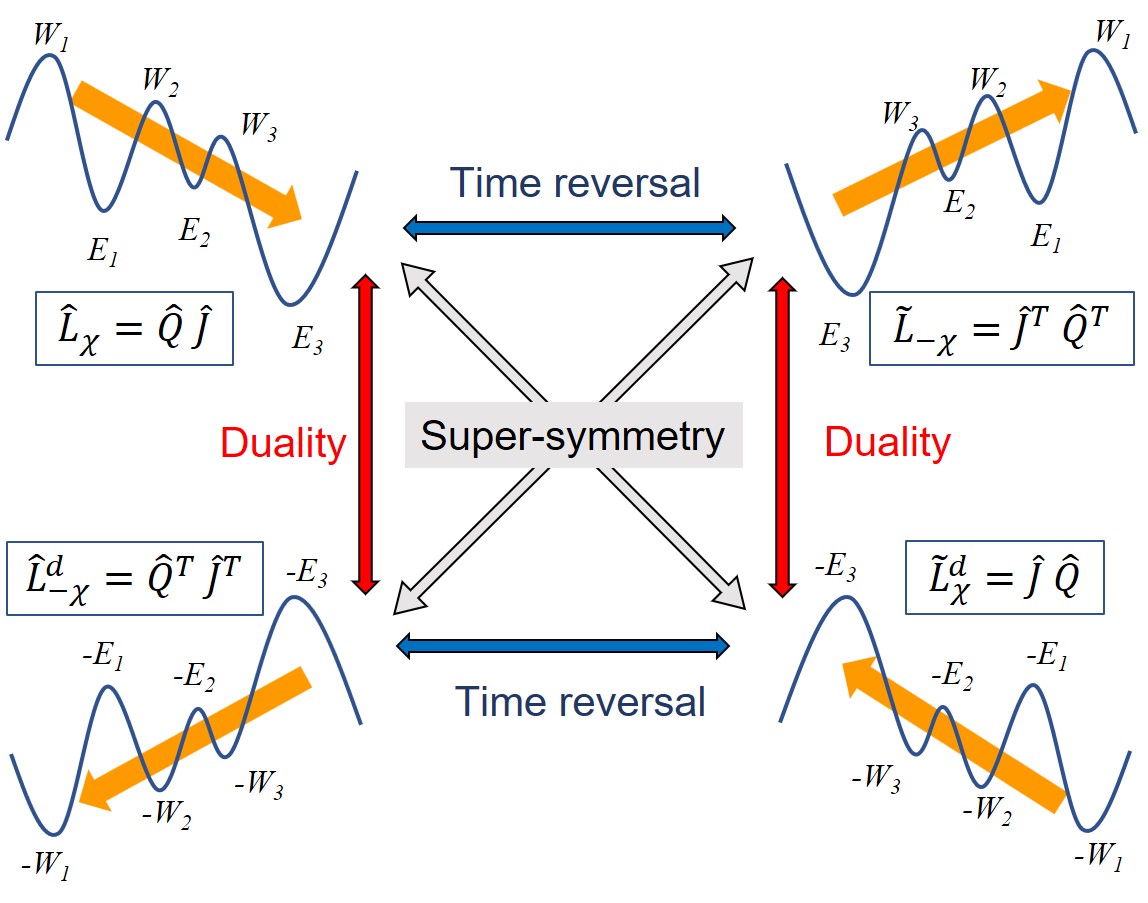}
 \caption{Symmetries in stochastic pump effect. The pumped systems is described by an operator $\hat{L}_\chi$, with the auxiliary parameter $\chi$ playing the role of counting field for the stochastic flow. The operators are composed of the current operator $\hat{J}_\chi$ and the divergence operator $\hat{Q}_\chi$( their dependence on the counting parameter $\chi$ suppressed). The symmetry relations between the four operators are shown. The superscript $d$ implies applying the duality transformation. The operators with hat/tilde above are related by the time-reversal symmetry. The superscript $T$ denotes taking the transposed operator. Here, the blue potential landscape and the orange current arrow schematically illustrate the action of various symmetries on a pumped Brownian systems. Figures are reproduced from \cite{ren2011duality}. }
 \label{fig7}
 \end{figure}

 The non-adiabatic pump regime promises fruitful future applications, as illustrated by the non-adiabatic control method. Therefore, a general conclusion valid across this vast regime would be significantly instructive. Importantly, general restrictions on stochastic pumps based on symmetry arguments are proposed~\cite{ren2011duality}. The intertwined implication of supersymmetry, duality, time-reversal symmetry is systematically discussed. Basically speaking, the duality interchanges the space of states/positions(potential local minima) and fluxes(potential barriers), and the time-reversal maps one time-dependent realization into another with the reversed driving protocol and the reversed stochastic trajectory. Meanwhile, the supersymmetry is a combination of the above two symmetries, as shown in Fig.~\ref{fig7}. Specifically, the master/Fokker-Planck operator in a stochastic dynamics can be split as the multiplication of two parts: the current operator $\hat{J}(\chi)$ and the divergence operator $\hat{Q}(\chi)$. Here $\chi$ is the counting parameter for the transported quantity forming the stochastic flow. The action of $\hat{J}(\chi)$ on a distribution evaluates the stochastic current flow, while the divergence $\hat{Q}(\chi)$ acting on the current flow leads to the changing rate of probability distribution. By identifying $\hat{L}(\chi)=\hat{Q}(\chi)\hat{J}(\chi)$ as the original system, we can obtain the other three derivative systems, by the transformed operators as shown in 
Fig.~\ref{fig7}, which are $\hat{L}^d=\hat{Q}^T(\chi)\hat{J}^T(\chi)$(duality), $\tilde{L}^d=\hat{J}(\chi)\hat{Q}(\chi)$(supersymmetry), and $\tilde{L}=\hat{J}^T(\chi)\hat{Q}^T(\chi)$(time reversal), respectively~\cite{ren2011duality}.

We note here that, the operator after time-reversal transformation defined here is closely related to the backward Fokker-Planck/Kolmogorov equation in the diffusive systems, so called adjoint operator~\cite{risken1996fokker}. 
An additional similarity transformation could be required in the form of $P_{st}\tilde{L}P^{-1}_{st}$, in order to make the connection with physically achievable systems as the time-reversal counterpart. Here $P_{st}$ is the steady state distribution of the original system\cite{kurchan1998fluctuation,harris2007fluctuation}. This similarity transformation is suppressed in the schematic Fig. \ref{fig7} to stay succinct. These symmetry transformations would provide us a wide range of exact consequences and convenient analysis. Here we mention some of the interesting ones. 
 
 Consider one potential landscape with a set of local meta-stable states. The kinetics is basically described by a set of potential well depth and barrier height. We sketch the action of symmetries in this example in Fig.~\ref{fig7}. If only the barrier is temporally modulated, there would be no net pumping current since the distribution keeps unchanged, as demanded by the unchanged local detailed balance condition. Using a duality mapping between wells and barriers, it is shown that wells are barriers are exchanged so that  the potential landscape is up-side-down. As such, the sole driving of potential wells is equivalent to the sole driving of barriers in the dual system, which can not provide net directional current even though the system is highly non-equilibrium~\cite{ren2011duality}. Therefore, simultaneous driving of the wells and barriers is a necessary condition for inducing non-vanishing net pumping. 
 
 In addition, symmetry similarly helps nail down the stochastic distribution properties. If the detailed fluctuation theorem(FT) $\frac{P_F(J)}{P_B(-J)}=e^{AJ}$ is verified in one system, it is simultaneously proved that its dual system has the FT given by $\frac{P_F^d(J)}{P_B^d(-J)}=e^{-AJ}$. Here, the subscripts $F$ and $B$ stands for the forward and backward driving protocols, and the superscript $d$ represents the duality transformation. $A$ is the thermodynamic affinity and $J$ is the stochastic flux. 
 
 The examples given here illustrate the power and elegance of symmetry arguments, since one theorem leads to a series of conclusions exactly true, regardless of the driving details. More related research following this logic in the future would be fruitful. 
 
\section{Conclusions and Perspectives}
\label{sec6}
In conclusion, the geometric heat pump effect pervades a whole range of dissipative open systems, for different underlying microscopic dynamics. As long as the coarse grained stochastic Markovian description is guaranteed, the separation of the pumped heat into the dynamical and geometric phase components is universally formulated. This theoretical framework has been utilized to study the molecular junction systems~\cite{ren2010berry}, the spin-boson systems~\cite{chen2013dynamic, wang2017unifying}, quantum Brownian systems~\cite{carrega2015functional}, the coupled chain systems~\cite{ren2012geometric, arrachea2012microscopic}, the opto-mechanical systems~\cite{nie2020berry}, etc. In spite of the vast scope of these studies and the far-reaching possible applications, there are still some perspective open questions to be addressed in the future research. Here we list some of them as the end of this review.

\begin{enumerate}
\item 
\emph{Searching for More Physical Platforms. }
The investigated setups are restricted to the two-terminal systems, in which there is often only one kind of carrier of the transported heat. It is possible that the incorporation of more than two terminals would provide us more bizarre phenomena of heat pump. One of the questions along this line is about the non-Hermitian heat pump~\cite{xu2016topological}. How would the introduction of the gain and loss channel affect the fluctuation properties of the pumped heat? Another question to be answered regards the influence of interaction between multiple degrees of freedom, like in the hybridized systems~\cite{wang2020managing}, thermal radiation systems~\cite{li2019adiabatic}, and multi-component systems~\cite{ren2014third}. Also, the geometric heat pump effect in the strong system-bath coupling regime\cite{talkner2020colloquium} is an interesting topic. These future studies would help us in proposing more functional thermal devices realizable in experiments. The experimental realization of the geometric heat pump is also a task to be fulfilled in future studies.

\item \emph{With Thermodynamic Uncertainty Relation and Non-equilibrium Response Theory. }
The geometric contribution can be utilized to construct heat engines~\cite{giri2017geometric, bhandari2020geometric, hino2020geometrical, brandner2020thermodynamic}.The geometric heat pump process would incur excess entropy production~\cite{sagawa2011geometrical} and this would leave signature in the power-efficiency relations~\cite{shiraishi2016universal, segal2008stochastic}. In view of the recently discovered series of thermodynamic uncertainty relations(TUR)~\cite{horowitz2020thermodynamic}, it would be important in discussing the restriction on the geometric heat pump exerted by the TUR. Also important is the connection of the geometric contribution in driven systems to their non-equilibrium response theory~\cite{sagawa2011geometrical, lippiello2014nonequilibrium}.

\item \emph{Non-adiabatic Control over Quantum Geometric Heat Pump. }
The non-adiabatic control paradigm has been successfully demonstrated in the classical stochastic pump systems~\cite{funo2020shortcuts, takahashi2020nonadiabatic}. In the non-adiabatically driven quantum systems, the effect of quantum coherence becomes significant~\cite{wang2017unifying}. The application of non-adiabatic control tools to quantum systems is not a straightforward task, but it would pay off in its application to the advancement of quantum technology. 

\end{enumerate}

\begin{acknowledgments}
Z.W., L.W., J.C., and J.R. acknowledge the support by the National Natural Science Foundation of China (No. 11935010 and No. 11775159), Natural Science Foundation of Shanghai (No. 18ZR1442800 and No. 18JC1410900). C.W. acknowledge support from the National Natural Science Foundation of China (Grant No. 11704093) and the Opening Project of Shanghai Key Laboratory of Special Artificial Microstructure Materials and Technology. 
\end{acknowledgments}

\begin{appendix}

\section{The General Twisted Master Equation}
\label{appendix_quantum}
The total Hamiltonian of an open system in contact with two heat baths can be divided into $\hat{H}=\hat{H}_S+\hat{H}_L+\hat{H}_R+\hat{H}_{SL}+\hat{H}_{SR}$, where the subscripts $S$, $L$ and $R$ denote the middle system, the left bath and the right bath respectively. The interaction part $\hat{H}_{SL}+\hat{H}_{SR}$ has a tensor product form: $\hat{H}_{SL}=\hat{a}_1\bigotimes \hat{b}_1$ and $\hat{H}_{SR}=\hat{a}_2\bigotimes \hat{b}_2$. Operators $\hat{a}$'s act on the system and $\hat{b}$'s act on the baths. This model describes the heat transport process of a nano-scale system embedded in macroscopic heat baths. The transported heat in a given time duration is usually much larger than the energy of the middle system and its interaction with heat baths. Under this assumption, the FCS of the transported heat from the left heat bath to the system can be well described by the CGF of the change of $\hat{H}_L$. The two-point measurement technique is well suited in the calculation~\cite{esposito2009nonequilibrium}.

$\hat{H}_L$ has a set of eigenvalues and eigenvectors given by $\hat{H}_L\ket{n}=E_n \ket{n}$. Two measurements of the left bath energy are carried out at time $0$ and $t$. The probability of their being $m$ and $n$ is given by the measurement postulate of quantum mechanics:
\begin{equation}
P(n,t;m,0)=\braket{n| \hat{U}(t,0)|m} \braket{m | \hat\rho_0 | m} \braket{m | \hat{U}^\dagger (t,0) | n},  
\end{equation}
where $\hat{\rho}_0$ is the initial density matrix evaluated at time $0$. The operator $\hat{U}(t,0)$ is the propagator from time $0$ to time $t$, with the defining property $i \frac{\partial}{\partial t}\hat{U}(t,0)=\hat{H}(t)\hat{U}(t,0)$. The heat entering the interested bath is equal to the energy difference $E_n-E_m$. Therefore, the characteristic function of heat is
\begin{equation}
\begin{split}
    Z(\chi, t)&\equiv\braket{e^{\chi(E_n-E_m)}}
    = \sum_{m,n}e^{\chi(E_n-E_m)}P(n,t;m,0)
    \\
    &=\textrm{tr} \left \{ \tilde{U}(\chi, t) \tilde{\rho}_0 \tilde{U}^\dagger(\chi, t) \right \}, 
\end{split}
\end{equation}
where the angle bracket denotes evaluating the average, 
\begin{equation}
\tilde{U}(\chi, t)=e^{\chi \hat{H}_L /2}\hat{U}(t,0) e^{-\chi \hat{H}_L /2}
\end{equation}
is the twisted propagator, $\textrm{tr}\{\hat{A}\}$ means calculating the trace of an arbitrary operator $\hat{A}$, and $\tilde{\rho}_0=\sum_{m}\ket{m} \braket{m|\hat{\rho}_0|m} \bra{m}$ is the initial density matrix projected along the energy eigenstates of the baths $\{ \ket{m} \}$. This projection is automatically done if the baths are initially thermalized, since its density matrix is diagonal in the energy basis. We can define $\tilde{\rho}(\chi, t)=\tilde{U}(\chi, t)\tilde{\rho}_0 \tilde{U}^\dagger (\chi, t)$ as the density matrix with auxiliary counting field in the Schr{\"o}dinger picture, so that $Z(\chi,t)=\textrm{tr}\{\tilde{\rho}(\chi,t)\}$. Using these elegant definitions, the similarity between the quantum and classical formulation is manifest. Back to the interaction picture for later convenience, $\hat{\rho}^I(\chi, t)=e^{i(H_S+H_L+H_R)t} \tilde{\rho}(\chi, t) e^{-i(H_S+H_L+H_R)t}$ is the twisted density matrix, where the superscript $I$ means being in the interaction picture.

The twisted Liouville equation for $\hat{\rho}^I(\chi, t)$ is
\begin{equation}
\begin{split}
\frac{\partial}{\partial t}\hat{\rho}^I(\chi, t)=-i\{ (\hat{H}_{SR}^I +e^{\chi \hat{H}_L^I/2}\hat{H}_{SL}^I e^{-\chi \hat{H}_L^I/2} )\hat{\rho}^I(\chi, t) \\
-\hat{\rho}^I(\chi, t)(\hat{H}_{SR}^I+e^{-\chi \hat{H}_L^I/2} \hat{H}_{SL}^I e^{\chi \hat{H}_L^I/2})            \}.
\end{split}
\end{equation}
This integro-differential equation can be transformed into a non-Markovian master equation by firstly using the Born approximation and then taking trace over the bath variables. If the weak coupling and large thermalized reservoir limit are assumed, the total density matrix can be written in a tensor product form $\hat{\rho}^I(\chi, t)=\hat{\rho}_S^I(\chi,t) \bigotimes \hat{\rho}_L^I\bigotimes \hat{\rho}_R^I $. The density matrices of two reservoirs can be taken to be of the Gibbs form. Averages $\left \langle \hat{b}_1 \right \rangle=\left \langle \hat{b}_2 \right \rangle=0$. If additional time scale separation between the baths and the system is assumed, the Markovianity is ensured~\cite{gardiner2004quantum}. The twisted Markovian master equation has the general form:
\begin{equation}
\begin{split}
    & \frac{\partial}{\partial t}\hat{\rho}_S^I(t)=\\
    & \int_{0}^{\infty}dt' \{ \hat{a}_2^I(t-t')\hat{\rho}_S^I(t)\hat{a}_2^I(t) \left \langle \hat{b}_2^I(t)\hat{b}_2^I(t-t') \right \rangle+ \\
     & \hat{a}_1^I(t-t')\hat{\rho}_S^I(t)\hat{a}_1^I(t)\left \langle e^{-\chi \hat{H}_L/2}\hat{b}_1^I(t)e^{\chi \hat{H}_L}\hat{b}_1^I(t-t')e^{-\chi \hat{H}_L/2} \right \rangle\\
     & +\hat{a}_2^I(t)\hat{\rho}_S^I(t)\hat{a}_2^I(t-t')\left \langle \hat{b}_2^I(t-t')\hat{b}_2^I(t) \right \rangle+ \\
     & \hat{a}_1^I(t)\hat{\rho}_S^I(t)\hat{a}_1^I(t-t')\left \langle e^{-\chi \hat{H}_L/2}\hat{b}_1^I(t-t')e^{\chi \hat{H}_L}\hat{b}_1^I(t)e^{-\chi \hat{H}_L/2} \right \rangle \\
     & -\hat{a}_2^I(t)\hat{a}_2^I(t-t')\hat{\rho}_S^I(t)\left \langle \hat{b}_2^I(t)\hat{b}_2^I(t-t') \right \rangle -\\
     & \hat{a}_1^I(t)\hat{a}_1^I(t-t')\hat{\rho}_S^I(t)\left \langle e^{\chi \hat{H}_L^I/2}\hat{b}_1^I(t)\hat{b}_1^I(t-t')e^{-\chi \hat{H}_L/2} \right \rangle \\
     & -\hat{\rho}_S^I(t)\hat{a}_2^I(t-t')\hat{a}_2^I(t)\left \langle \hat{b}_2^I(t-t')\hat{b}_2^I(t) \right \rangle- \\
     & \hat{\rho}_S^I(t)\hat{a}_1^I(t-t')\hat{a}_1^I(t)\left \langle e^{-\chi \hat{H}_L/2}\hat{b}_1^I(t-t')\hat{b}_1^I(t)e^{\chi \hat{H}_L/2} \right \rangle  \}
\end{split}
\end{equation}
The angle brackets denotes taking the ensemble average. Setting the counting field $\chi=0$, we can recover the trace-preserving master equation. Either by going into the super-operator formalism or by ignoring the coherence effect and adopting the twisted Pauli master equation, the twisted master equation can be simplified to
\begin{equation}
\label{twiste_master}
\frac{\partial}{\partial t}\ket{\rho(\chi, t)}=\hat{L}(\chi, t)\ket{\rho(\chi, t)}.
\end{equation}
Here the vector $\ket{\rho(\chi,t)}$ is the vector representation of the twisted density matrix. $\ket{\rho(\chi,t=0)}|_{\chi=0}$ is identified with the vector form of the matrix $\textrm{tr}_L\textrm{tr}_R\{\tilde{\rho}_0\}$, where $\textrm{tr}_L\textrm{tr}_R$ means tracing over all baths variables. The characteristic function of the conducted heat is, formally analogous to the classical case, $Z(\chi, t)=\braket{-| \rho(\chi,t)}$.

\section{The General Twisted Fokker-Planck Equation}
\label{appendix_classical}
Here we assume the stochastic characteristics of the microscopic dynamics of coupled oscillators. This phenomenological approach is powerful both in doing numerical calculations and deriving analytical results \cite{harris2007fluctuation}. Specifically, the time evolution of the reduced degrees of freedom of the middle system is described by the following Langevin equation:
\begin{equation}
\begin{split}
    & \dot{x}_n=p_n=\frac{\partial H}{\partial p_n}, \\
     & \dot{p}_n=-\frac{\partial H}{\partial x_n}+(-\Gamma_1 p_1+\xi_1)\delta_{n1}+(-\Gamma_N p_N+\xi_N)\delta_{nN},
\end{split}
\end{equation}
where $H$ is the Hamiltonian of the middle system. Two heat baths, acting on the $1$-st particle and $N$-th particles of the middle system respectively, are of the Langevin form, with the noise being Gaussian and white. $\left \langle \xi_i (t)\right \rangle=0$ and $\left \langle \xi_i (t_1)\xi_j (t_2) \right \rangle=2\Gamma_i T_i \delta_{ij} \delta(t_1-t_2)$. In the above equations, $\delta_{ij}$ and $\delta(t_1-t_2)$ are Kronecker delta and Dirac delta respectively. The dynamical equations describe an open classical systems, coupled to the left and right thermal baths via the $1$st and $N$th particle respectively, with other degrees of freedom only directly modulated by the conservative Hamiltonian $H$. Additionally, the total transported heat $Q_1$ entering from the left bath into the middle system accumulates following $\dot{Q}_1=p_1(-\Gamma_1 p_1 +\xi_1)$. The definition of $\dot{Q}_1$ can be understood as the fluctuating power exerted by the left bath, which is the multiplication of the velocity and the force\cite{sekimoto2007microscopic}. We note that, in the long time limit, the heat leaving the left bath should be equal to the that entering the right bath. This is due to the first law of thermodynamics and the energy of the middle system being bounded. Also the statistics of $Q_1$ and $Q_N$ approaches each other in the long time limit. We no longer discriminate between them and denote them as $Q$ in the following discussion. The joint distribution $P(x,Q,t)$ is totally determined by the initial condition and the generalized Fokker-Planck equation $\frac{\partial}{\partial t}P(x,p,Q,t)=\hat{L}_0P(x,p,Q,t)$. The operator $\hat{L}_0$ can be derived using the method of stochastic Liouville equation~\cite{kubo2012statistical}. In our interested classical system connected to two thermal baths, the operator $\hat{L}_0$ is
\begin{equation}
\begin{split}
    & \hat{L}_0=\left \{H, \right \}+\sum_{i=1/N}( \Gamma_i T_i\frac{\partial^2}{\partial p_n^2}+\Gamma_n \frac{\partial}{\partial p_n}p_n ) \\
    & +\Gamma_1(T_1+p_1^2)\frac{\partial}{\partial Q}+2p_1\Gamma_1T_1\frac{\partial^2}{\partial Q \partial p_1}+\Gamma_1 T_1 p_1^2\frac{\partial ^2}{\partial Q^2}, 
\end{split}
\end{equation}
where $\left \{H, \right \}$ is the Poisson bracket. The summation is over the $1$-st and $N$-th particles, which are directly connected to the baths. 

By defining $\rho(x,p,\chi,t)=\int_{-\infty} ^\infty dQ P(x,p,Q,t)e^{\chi Q}$, the characteristic function of $Q$ can be written as
\begin{equation}
Z(\chi,t)\equiv \braket{e^{\chi Q}}=\int_{-\infty}^{\infty} dx dp \rho(x,p,\chi,t).
\end{equation}
Here, the angle bracket indicates taking the average. Thus, calculating $\rho(x,p,\chi,t)$ is sufficient to derive the full CGF of transported heat in the given time period. The twisted Fokker-Planck equation $\frac{\partial}{\partial t}\rho(x,p,\chi,t)=\hat{L}\rho(x,p,\chi,t)$ does exactly this task. $\hat{L}=e^{\chi Q}\hat{L}_0e^{-\chi Q}$. Formally, it is related to $\hat{L}_0$ via the substitution $\frac{\partial}{\partial Q}\rightarrow -\chi$. $\hat{L}(\chi)$ has the explicit form of
\begin{equation}
\begin{split}
    & \hat{L}=\left \{ H,  \right \}+ \sum_{i=1/N}\left [\Gamma_i T_i \frac{\partial ^2}{\partial p_i^2}+\Gamma_i \frac{\partial}{\partial p_i}p_i\right ]\\
     & -2\Gamma_1 T_1\chi p_1\frac{\partial}{\partial p_1}-\Gamma_1 T_1 \chi +\Gamma_1 p_1^2 \chi(\chi T_1-1).
\end{split}
\end{equation}
The first part contains the Poisson bracket, generating the deterministic dynamics of the middle system. The stochastic influence of the heat baths enters the dynamics via the latter additional terms. 

\section{Geometric and Dynamic Phases in Floquet Systems}
\label{appendix_geo}
Consider a linear evolution equation
\begin{equation}
\label{app 1}
\frac{\partial}{\partial t}\ket{\psi(t)}=\hat{L} \ket{\psi(t)},
\end{equation}
where time-dependent $\hat{L}(t)$ changes cyclically in a time period $T_p$ under external driving. According to the Floquet theory, its solution can be written in the form of
\begin{equation}
\label{app 2}
\ket{\psi(t)}=e^{G(t)}\ket{\phi(t)}, 
\end{equation}
where $\ket{\phi(t+T_p)}=\ket{\phi(t)}$ is a cyclic state. This ansatz implies that the Floquet state only accumulates a phase-like factor $G(T_p)$ (thinking about Eq.~\ref{app 1} as the Schr\"odinger equation in the imaginary time domain) relative to the initial state when one driving period is accomplished. We define a cyclic bra $\bra{\tilde{\phi}(t)}$ with property $\frac{\partial}{\partial t}\braket{\tilde{\phi}(t)|\phi(t)}=0$. By differentiating Eq.~\ref{app 2} with respect to $t$ and taking the inner product of both sides with $\bra{\tilde{\phi}(t)}$, we arrive at the equation for $\lambda(t)$
\begin{equation}
\frac{d G(t)}{dt}=-\frac{\braket{\tilde{\phi}(t) |\frac{\partial}{\partial t}\phi(t)}}{\braket{\tilde{\phi}| \phi}}+\frac{\braket{\tilde{\phi}(t)| \hat{L}(t)| \phi(t)}}{\braket{\tilde{\phi}| \phi}}. 
\end{equation}

Therefore, following the spirit of Aharonov-Anandan(AA) phase~\cite{aharonov1987phase}, the total phase change across each period can be split into the dynamic and geometric contributions $G(T_p)=G_{\textrm{dyn}}+G_{\textrm{geo}}$ with the two contributions given by
\begin{equation}
\begin{split}
    & G_{\textrm{dyn}}=(\braket{\tilde{\phi}| \phi})^{-1}\int_0 ^{T_p} dt {\braket{\tilde{\phi}(t) | \hat{L}(t) | \phi(t)}}, \\
    & G_{\textrm{geo}}=-(\braket{\tilde{\phi}| \phi})^{-1}\int_0 ^{T_p} dt {\braket{\tilde{\phi}(t) | \frac{\partial}{\partial t} | \phi(t)}}. 
\end{split}
\end{equation}

\end{appendix}
\bibliography{ref.bib}

\onecolumngrid

\end{document}